# Introduction to Management Information system

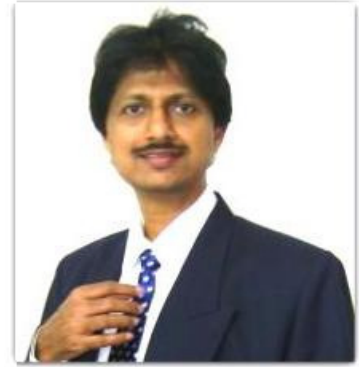

**By- Umakant Mishra, Bangalore, India**

umakant@trizsite.tk, http://umakant.trizsite.tk

**Contents**

## 1. What is MIS

A Management Information System (MIS) is a systematic organization and presentation of information that is generally required by the management of an organization for taking better decisions for the organization. The MIS data may be derived from various units of the organization or from other sources. However it is very difficult to say the exact structure of MIS as the structure and goals of different types of organizations are different. Hence both the data and structure of MIS is dependent on the type of organization and often customized to the specific requirement of the management.

The meaning of MIS is well represented by the three it consists. "Management" – the information system is built for management and not for the operational staff. "Information System"- an information system that takes care of sourcing, organizing and managing the required data and presenting in the desired formats that may be useful in a context and for a purpose.

## 2. The Source of MIS data

MIS data is generally summarized from the day-to-day operational data of the organization. Most part of the MIS database is collected from different sub-systems of the organization. The subsystems may be Human Resource System, Production Management System, Finance System, Sales Management System, Project Management System etc.

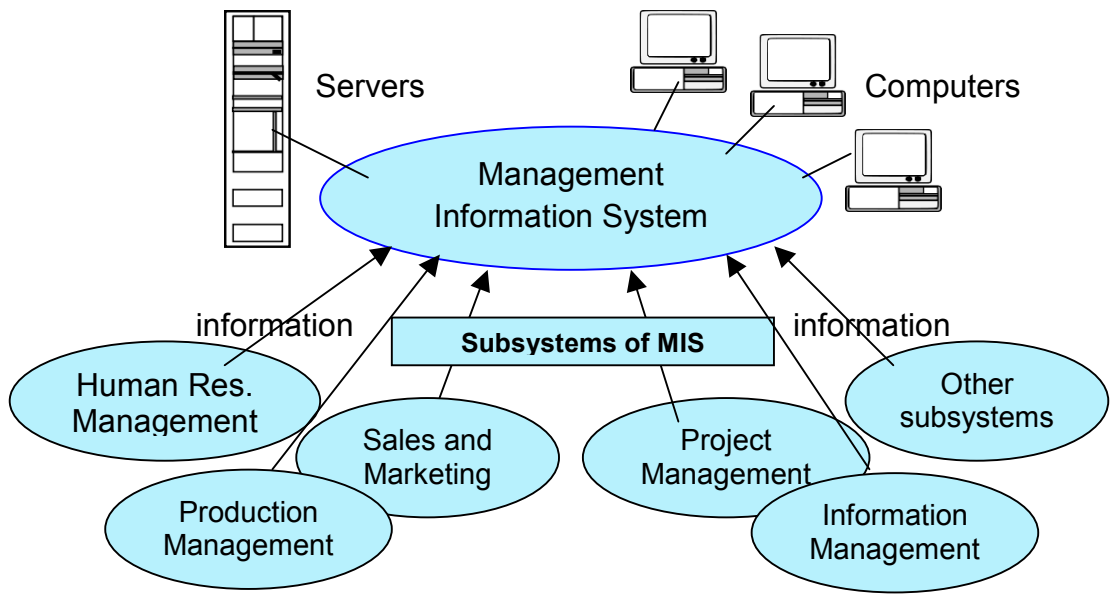

The above illustration shows how the critical ==management information== flows from every unit of organization to a centralized management information system. It will be interesting to note that there can be some "management information" from the "information management unit" to the "management information system" too[#].

Some types of MIS data are also collected from sources external to the organization, such as competitors' data, which may be obtained from different websites in the Internet.

## 3. Difference between MIS and "Management of Information System"

"Management of Information System" is same as "Information System Management". This refers a system of managing information. Although by definition an Information System is not necessary to adopt computerize the data, with the advancement of computing industry and popularity of computers, computerization has become an integral part of any information management system.

"Management Information System" is different from "Management of Information System" or "Information Systems Management". While "information management" can be for any purpose, MIS is meant for only the higher level managers. Hence in once sense MIS is a part or subset of an organizational information system.

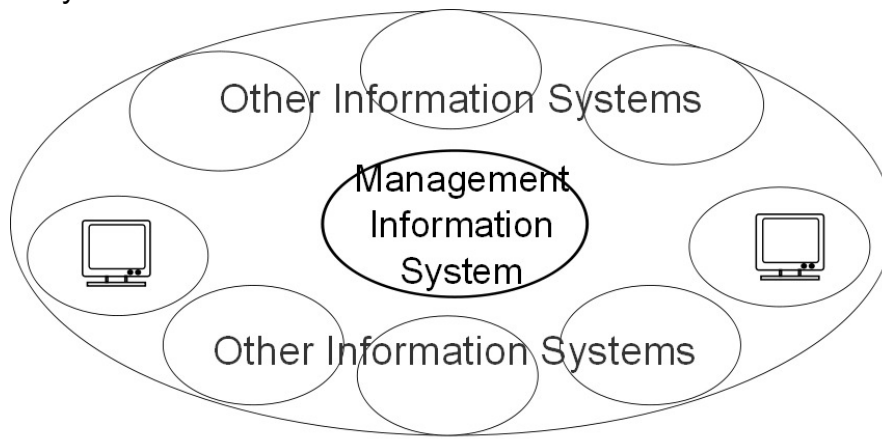

The information system of an organization

---

[#] Umakant Mishra, Introduction to Management Information System, available at http://ssrn.com/abstract=2307474

The information system of an organization may contain other information systems like financial information system (such as payroll etc.), sales information system (such as daily sales) etc. In those sense regular financial transactions, sales transactions etc. are not MIS. They fall into the category of "transaction processing system". We will discuss the difference between MIS and other information systems in a separate article.

## 4. Similarity between MIS and other information systems[#]

From technological aspect MIS is just another information system. It involves the same computing infrastructure and technologies as other information systems. The components of MIS, such as, input, output, data, computers, processing etc. are also the same as any other information systems.

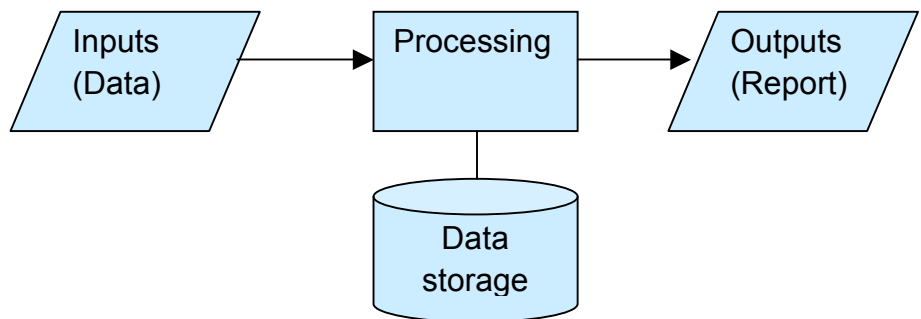

**Basic Information System Model**

- ⇨ The process of building an MIS involves similar steps of planning, analysis, design, development, testing and implementation etc. like any other information system.
- ⇨ The technology of MIS involves database, Internet, security and all such things that are used for other information systems.
- ⇨ In fact MIS is a part or subset of the Information Management System of an Organization as a whole.

---

1. [#] Umakant Mishra, *Management Information Systems Vs. Other Information Systems*, available at SSRN eLibrary, http://ssrn.com/abstract=2308846

# 5. Difference between MIS and other information Systems

In a previous articlewe had discussed how Management Information System (MIS) is different from Management of Information Systems. The main difference lies in their purpose of use. As an organization generally consists of various departments and involves various types of functions, a single type of information system cannot fulfill the needs of all these departments and functions. Hence different types of information systems are designed for different purposes.

For example the transaction processing system deals with the automation of day-to-day activities of the organization, such as, sales and marketing, manufacturing and production, finance and accounting etc. On the other hand the MIS brings the processed data periodically from various departments of the organization. This process increases the responsiveness of the management and helps the management in better planning, organizing, leading and controlling to increase productivity and bring competitive advantage to the organization.

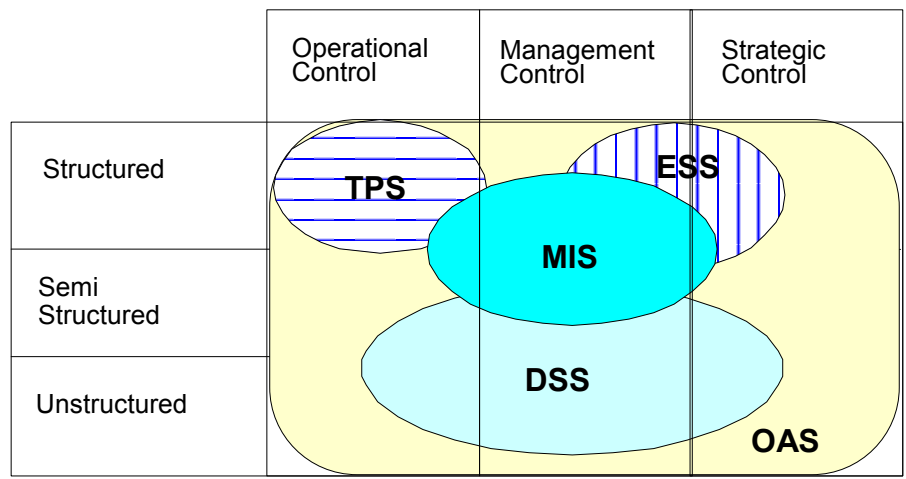

**Domains of Information System Types[#]**

The above diagram shows the domain of different types of Information Systems. While a transaction processing system is highly structured an MIS is not fully structured. The decision support systems range from semi structured to unstructured territory. Let's see the differences between MIS and some other types of information systems.

---

[#] Kroenk and Hatch, Management Information Systems, 3rd edition, pp 143, figure 4-3

## 6. MIS and Transaction Processing systems

Transaction processing systems (TPS) deal with computerization of any type of transactions such as financial transactions (receipts, payments etc.), sales transactions (purchases, sales etc.). The purpose of TPS is:

- ⇨ To track the routine activities and automate the day-to-day operations of specific activities of an organization such as sales, production, payroll etc.
- ⇨ To computerize day-to-day transactions like sales etc., print sales vouchers, prepare daily reports, summary reports etc. This system intends to reduce manual work and speed up operations by using computers.

At the operational level the tasks and goals are predefined and the processes are highly structured. There are specific rules on how a payslip is to be calculated and specific format on how a payslip is to be printed. Hence, in most cases, ready made software packages are available for transaction processing systems.

**Difference between MIS and TPS:**

- ⇨ While transaction processing systems are meant for minute to minute activities, the Management Information systems are meant for medium term planning and strategic decision making.
- ⇨ If a transaction processing system halts for an hour or a day it will affect the activities of the organization to incur loss. But the MIS data is not computed so frequently.
- ⇨ Transaction processing systems generally deal with high volume of data. The MIS typically deals with much less volume of data.
- ⇨ The transaction processing system is meant for operators and supervisors involved in specific operations. The MIS is meant for middle and higher level managers.
- ⇨ Often the TPS is used as input to MIS. The summary of TPS is periodically transferred to MIS.

## 7. Management Information Systems and Decision Support Systems

Decision support systems are meant for assisting the decision makers (human beings) in taking non-standard and complex decisions. Decision support systems use intelligence like human beings to decide "what if this happens…". Decision support systems are intelligent systems and similar to expert systems.

- ⇨ Decision Support Systems use data from both TPS and MIS. They also take information input from other sources such as competitors' reports, information from websites.
- ⇨ Decision support systems use various scientific methods and statistical calculations to arrive at better decisions on a given condition.
- ⇨ DSS do not necessarily follow conventional rules. For example, a conventional system can compute the price of a product based on the cost of production, but it cannot compute the price of product to be less than the cost of production. But sometimes a decision may be taken to sell the product at a price lower than the manufacturing cost (thereby incurring loss) because of competitor's price.
- ⇨ The decision support systems are based on highly unstructured data. As the problem environment is frequently changing there is no ready-made decision support software. There are various software tools for advanced statistical calculations and presentations. The decision maker may use some of these tools to fine-tune his decisions.

## 8. Management Information Systems and Executive Support Systems

Executive Support Systems are intended to provide necessary information to Executives and senior management to take non-routine and strategic decisions. ESS may get input from TPS of an organization and from other sources and present the information in formats that may be useful to the senior management to take effective decisions.

Thus as we saw above all these systems including MIS, DSS, ESS etc. are used by the senior management to take effective decisions. But there are certain differences between them so far as their field of application is concerned.

## 9. Management Information Systems and Office Automation Systems

Office automation systems use computerized systems to automate any workflow of the organization. For example, a manual payroll system may be computerized for an automatic calculation of salaries and other payments. Standard acknowledgements may be automatically sent by emails. The replies to queries may be standardized and to some extent automated.

- ⇨ There are many office automation systems such as MS Office Suite, Star Office (from Sun Microsystems), IBM Lotus symphony, Open Office etc.
- ⇨ Office automation systems have nothing to do with MIS. But they are used for all types of purposes at every level. Hence sometimes office automation tools are useful for compiling and presenting MIS reports/ presentations.

## 10. Purview and characteristics of MIS

However, the extent or purview of MIS may vary from organization to organization. A small organization cannot have the same extent of MIS as a large organization. However, a small organization having ten people still needs an MIS consisting of monthly production figures, purchase and sales figures, income and expenditures etc. as much as a large organization needs it. However, the mechanism of managing the MIS may differ from organization to organization. In a ten people organization the manager himself may maintain the MIS without a specialized MIS application. In a hundred people organization there may be a specialized MIS application that is maintained by the IT department. In contrast a thousand people large organization may have an MIS that is linked to the operational data of every unit (or cost center or profit center or regional office) to process and transfer the summary information to build the MIS database automatically. The characteristics of MIS is as follows.

- ⇨ MIS is linked to a business and intended for the top-level management of the organization. It is different from the operational information which deals with day-to-day activities of a business and different from personal information which is meant for personal use of an individual.
- ⇨ The structure of MIS depends on the structure and procedures of the organization. Every organization may not have the same concept of sales, clients or projects. The MIS system is dependant on the nature of organization or business. As the projects, activities and targets can be different for different organizations the structure and output of MIS can vary accordingly.
- ⇨ Generally the volume of data in MIS is smaller compared to other information systems in the organization. The inputs of MIS data is obtained from different subsystems or departments of the organization.

## 11. Levels of management and MIS

The lower level management takes control of the day-to-day operational activities. They follow organizational procedures and guidelines to achieve the quantitative targets. Middle level management takes tactical decisions with a medium term perspective, such as, for six months to two years. The middle level management is often associated with specific units or specific projects of the organization. The middle level managers such as departmental heads or project managers are supposed to take decisions for their department or project and not for the organization as whole.

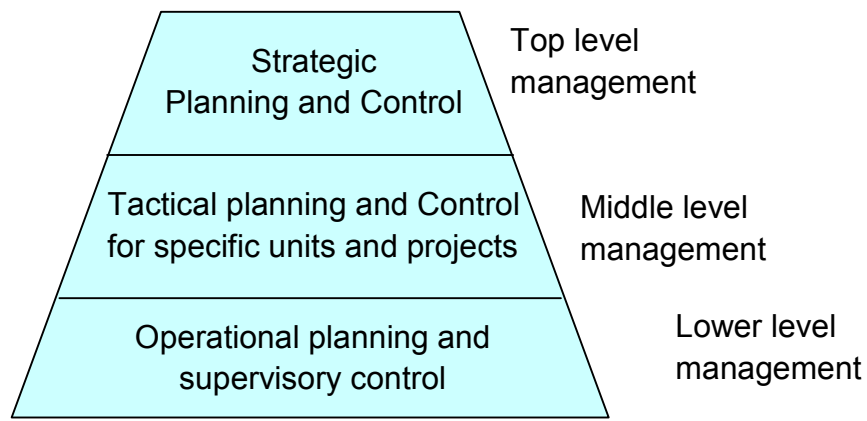

The top-level management takes strategic decisions for long terms such as for five years and more. As the levels of management and their objectives and deliverables are different, the information systems involved with different levels of management are different. While the lower level managers take decisions based on the summary reports of transaction processing systems, the middle level managers require MIS to take standard and nonstandard decisions. The higher level management requires MIS for standard and non-standard decisions and use other tools like DSS to take more effective strategic decisions.

## 12. Benefits of MIS

Information is key to decision making. As a manager is necessary for managing a business or organization, an MIS is necessary for the manager. The managers can take right decisions only if the appropriate information is available to them. If the critical information is not available or available in a half hazard or confusing format then the manager may fail to take an appropriate decision at the right time. Thus a well-organized MIS has the following benefits.

- ⇨ Facilitate easy access of critical/ regularly used data to the people in management. As the required information is well organized and quickly available to the management it increases the decision making potential of the management.

- ⇨ Quick availability of up-to-date information facilitates better planning, monitoring, and communication. It not only helps taking quick short-term decisions (because of quick availability of critical data), but also helps taking long-term decisions and strategic planning.

- ⇨ Helps standardization- As the MIS is extracted from different divisions of the organization, any incompatibility that may be found in different formats/procedures are abstracted. Any incompatibility of lower level operational data is aggregated and standardized in MIS.

- ⇨ A good MIS obviously improves management efficiency, operational efficiency, employee productivity, project performance and customers'/donors' satisfaction. A more effective management yields better competitive advantage.

## 13. Summary

A Management Information System (MIS) is a systematic organization and presentation of information that is generally required by the management of an organization. In one sense MIS is a part or subset of the Information Management System of an Organization. There are different types of information systems such as, Transaction Processing System (TPS), Decision Support System, Executive Support System (ESS) etc. having some differences and some overlapping.

MIS is different from DSS as the later uses unstructured data to take non-standard decisions. MIS is different from TPS and the later is used for day-to-day transactions of the organization. Thus the purpose of MIS is to organize and present such information that would help the middle and higher level management for taking better and more effective decisions to increase the productivity of the organization and gain competitive advantage.